\documentclass{aastex62}

\graphicspath{{./figures/}}

\newcommand{\ie}{{\it i.e.}}

\newcommand{\eg}{{\it e.g.}}

\newcommand{\figu}[1]{Fig.~\ref{fig:#1}}

\newcommand{\bi}{\begin{itemize}}
\newcommand{\ei}{\end{itemize}}

\received{???}
\revised{???}
\accepted{???}
\submitjournal{ApJ}

\shorttitle{UHECRs \& neutrinos from LL-GRBs}
\shortauthors{D. Boncioli et al.}

\begin{document}

\title{On the common origin of cosmic rays across the ankle and diffuse neutrinos at the highest energies \\ from low-luminosity Gamma-Ray Bursts}

\correspondingauthor{Denise Boncioli}
\email{denise.boncioli@desy.de}

\author{Denise Boncioli}
\affil{Deutsches Elektronen-Synchrotron (DESY), Platanenallee 6, 15738 Zeuthen, Germany}
\nocollaboration

\author{Daniel Biehl}
\affil{Deutsches Elektronen-Synchrotron (DESY), Platanenallee 6, 15738 Zeuthen, Germany}
\nocollaboration

\author{Walter Winter}
\affil{Deutsches Elektronen-Synchrotron (DESY), Platanenallee 6, 15738 Zeuthen, Germany}
\nocollaboration






\begin{abstract}
We demonstrate that the UHECRs produced in the nuclear cascade in the jet of Low-Luminosity Gamma-Ray Bursts (LL-GRBs) can describe the UHECR spectrum and composition and, at the same time, the diffuse neutrino flux at the highest energies. The radiation density in the source simultaneously controls the neutrino production and the development of the nuclear cascade, leading to a flux of nucleons and light nuclei describing even the cosmic-ray ankle at $5 \cdot 10^{18} \, \mathrm{eV}$. The derived source parameters are consistent with population studies, indicating a baryonic loading factor of about ten. Our results motivate the continued experimental search of LL-GRBs as a unique GRB population.
\end{abstract}



\section{Introduction} \label{sec:intro}

Gamma-Ray Bursts (GRBs) are extreme electromagnetic outbursts, see, for example, \citet{Piran:2004ba}. Here we consider the possibility that low-luminosity GRBs (LL-GRBs, with isotropic luminosity $\lesssim 10^{49} \mathrm{erg\,s^{-1}}$) and high-luminosity GRBs (HL-GRBs, with isotropic luminosity $\gtrsim 10^{49} \mathrm{erg\,s^{-1}}$) are two distinct populations, based on the different local rate of the two samples \citep{Guetta:2006gq,Liang:2006ci,Virgili:2008gp,Sun:2015bda}. Being locally much more abundant than HL-GRBs ($\approx 1 \, \mathrm{Gpc^{-3}\,yr^{-1}}$), LL-GRBs ($\approx 300 \, \mathrm{Gpc^{-3}\,yr^{-1}}$, as predicted in \citet{Liang:2006ci}) have been proposed as sources of cosmic rays and neutrinos \citep{Murase:2006mm,Murase:2008mr,Liu:2011cua,Senno:2015tsn}. More recently, LL-GRBs as sources of ultra-high energy cosmic-ray (UHECR) nuclei have been studied in \citet{Zhang:2017moz} including possible injection compositions. Due to the low radiation density, it has been proposed that the nuclei can escape intact from the sources, leading to compatibility with the UHE chemical composition measured by the Pierre Auger Observatory \citep{Aab:2014kda} after propagation. However, the low radiation density required for nuclei to escape implies at the same time low neutrino production efficiencies -- possibly too low to simultaneously describe the diffuse neutrino flux in a one zone model.

HL-GRBs have been tested as the possible origin of UHECRs for both protons in \citet{Baerwald:2014zga} and nuclei in \citet{Biehl:2017zlw} describing cosmic-ray and neutrino data explicitly. It has been shown that for nuclei and for high enough radiation densities, a nuclear cascade due to the photo-disintegration of nuclei develops -- while at the same time  neutrinos are efficiently produced by photo-hadronic interactions. Very tight constraints on neutrinos from HL-GRBs have been obtained by using direction, timing and energy information from GRB catalogues for stacking limits \citep{Abbasi:2012zw,Aartsen:2017wea}. These constraints limit the parameter space to low radiation densities, such as high collision radii and low luminosities in the internal shock model -- parameters which may not be favorable for HL-GRBs, and point already towards LL-GRBs \citep{Biehl:2017zlw}. A possible caveat are multi-zone collision models in which the different messengers originate from different regions of the same GRB, predicting somewhat lower neutrino fluxes~\citep{Bustamante:2014oka,Globus:2014fka,Bustamante:2016wpu} -- which however cannot explain the diffuse neutrino flux. The stacking bounds do not apply to LL-GRBs due to their much longer duration (making the background suppression less efficient) and their low luminosity (limiting the detection of resolved sources). Note that the luminosity mentioned here represents the X-ray luminosity, which may differ from the intrinsic kinetic luminosity of the jet. The latter can be higher by a factor $\sim 100$ taking into account the energy conversion efficiency \citep{Aloy:2018czj}.

In this work, we study if LL-GRBs with a nuclear cascade in the jet can power the diffuse neutrino and cosmic-ray fluxes  at the highest energies at the same time, using methods similar to \citet{Biehl:2017zlw,Biehl:2017hnb}. We inject a nuclear composition which is found to be reasonable in the jet of GRB progenitors \citep{Woosley:2005gy,Zhang:2017moz}, and we include the transition to the next population (at lower energies). As an important ingredient, it was noted in \citet{Unger:2015laa} in a generic model and in \citet{Globus:2015xga,Biehl:2017zlw} for GRBs that the nuclear cascade also controls the production of nucleons below the change of the slope in the measured CR energy spectrum, called the ``ankle'' at $\sim 5 \cdot 10^{18} \, \mathrm{eV}$ \citep{Fenu:2017}, \ie, spectrum and composition may be described in a much larger energy range across the ankle. Our analysis is based on a combined source-propagation model, which means that we include the interactions of the injected nuclei in the source in addition to the UHECR propagation, whereas a propagation-only model starts off at the interface between source and extragalactic space. Compared to earlier studies, we perform extensive parameter space scans focusing on a combined description of UHECR and neutrino data and including the description of the cosmic-ray ankle. Similar to previous studies, we use the internal shock scenario as a baseline scenario and comment on alternatives where applicable. We also motivate future searches in next-generation telescopes such as CTA.

\section{Methods}\label{sec:methods}

Motivated by diffusive shock acceleration in the jet, we assume that the spectrum of the primary injected nuclei follows a power law $\propto E^{-2} \exp(-E/E_{\mathrm{max}})$. The maximal energy $E_{\mathrm{max}}$ is determined from the balance between acceleration and interaction rates, where we take into account adiabatic cooling, photo-hadronic interactions and synchrotron losses. The acceleration rate is given by $t'^{-1}_{\text{acc}} = \eta c/R'_L$ with the acceleration efficiency $\eta$ and the Larmor radius $R'_L = E'/ZeB'$ of a particle with energy $E'$ and charge number $Z$ (primed quantities refer to the shock rest frame); we use efficient acceleration $\eta\simeq 1$ in this study, which is degenerate with the injected composition and the energy scale uncertainties of Auger, as discussed in \citet{Biehl:2017hnb}. Our injection composition is a simplified version of the silicon-rich case defined in \citet{Zhang:2017moz} ($60\%$ ${}^{16}$O and $40\%$ ${}^{28}$Si into the jet).

The target photon field of the GRB prompt emission is modeled as a broken power law with lower and upper spectral index $\beta_1 = 1.0$ and $\beta_2 = 2.0$, respectively, and a break energy which is typically around $\varepsilon'_{\gamma, \mathrm{br}} \sim 1$ keV. We have tested that our results are not very sensitive to the exact values presented here. Accelerated nuclei interact with these target photons in internal shocks at a distance $R \simeq  2 \Gamma^2 c t_v$ from the engine, where $\Gamma$ represents the Lorentz factor and $t_v$ the variability time scale of the emission. In our calculations, we fix $\Gamma \simeq 10$ \citep{Aloy:2018czj} and vary the radius $R$ over a large range between $10^8$ km and $10^{12}$ km. The time variability thus changes from $10^0$ s to $10^4$ s. We assume a total duration of $\sim 2 \cdot 10^5$ s, which may be somewhat longer than the typical durations expected from observations ($10^3$ to $10^4 \, \mathrm{s}$). 
However, for GRBs with shorter durations (of the order of tens of seconds), the jet breakout time might be too large for the jet to be successfully launched \citep{Bromberg:2011fm}. Such choked jet sources are generally not expected to emit UHECRs, as cosmic rays strongly cool in the environment, possibly producing high energy neutrinos \citep{He:2018lwb}. Furthermore, note that our results will be degenerate in duration  times baryonic loading times apparent local rate, which means shorter durations can \eg\ be compensated by larger baroynic loadings. Changes in the Lorentz factor can be compensated by adjusting the radius or the variability time scale. The parameters chosen here are consistent with the ones used for jet formation and survival \citep{Aloy:2018czj}.

To simulate the nuclear interactions within the LL-GRB jet, we use the {\it NeuCosmA} code similar to \citet{Biehl:2017zlw,Biehl:2017hnb}, which is based on SOPHIA \citep{Mucke:1999yb} for photo-meson production (photon energy in nucleus' rest frame $\varepsilon_\gamma \gtrsim 150$ MeV). For photo-meson production off nuclei, a superposition model is used accordingly, \ie, the cross sections scale approximately with the nucleus' mass number $\sigma_{A\gamma} \approx A\sigma_{p\gamma}$. The photo-disintegration ($\varepsilon_\gamma \lesssim 150$ MeV) cross sections are taken from CRPropa 2 ($A < 12$) \citep{Kampert:2012fi} and TALYS ($A \geq 12$) \citep{Koning:2007}. For details on the interaction models, see \citet{Boncioli:2016lkt}.

A critical ingredient connecting the physics of source and propagation of UHECRs is the cosmic-ray escape mechanism from the source. One possibility for GRBs was proposed in \citet{Baerwald:2013pu}, postulating that, even in an expanding shell, the particles within the Larmor radius of the edge of the shell can escape. As a consequence, the escape rate scales $\propto R'_L \propto E'$, which means that the escaping spectra are one power harder than the spectra within the source (``direct escape''). A similar result is obtained for Bohm-like diffusion of particles escaping a compact region. If it is assumed that only the particles at the highest energies can escape, the spectrum may be even harder as in \citet{Globus:2014fka} or in \citet{Zhang:2017moz}, where the ejected spectra are defined as $\propto \exp(-\ln^2(E/E_{\text{max}}))$, as found in \citet{Ohira:2009rd}; we refer to this case as ``hard escape''. We use this choice in this study, since we have verified that this assumption is favored by the UHECR data with respect to the direct escape for the source evolution used for LL-GRBs.

\begin{figure}[t!]
\centering
\includegraphics[width=0.39\textwidth]{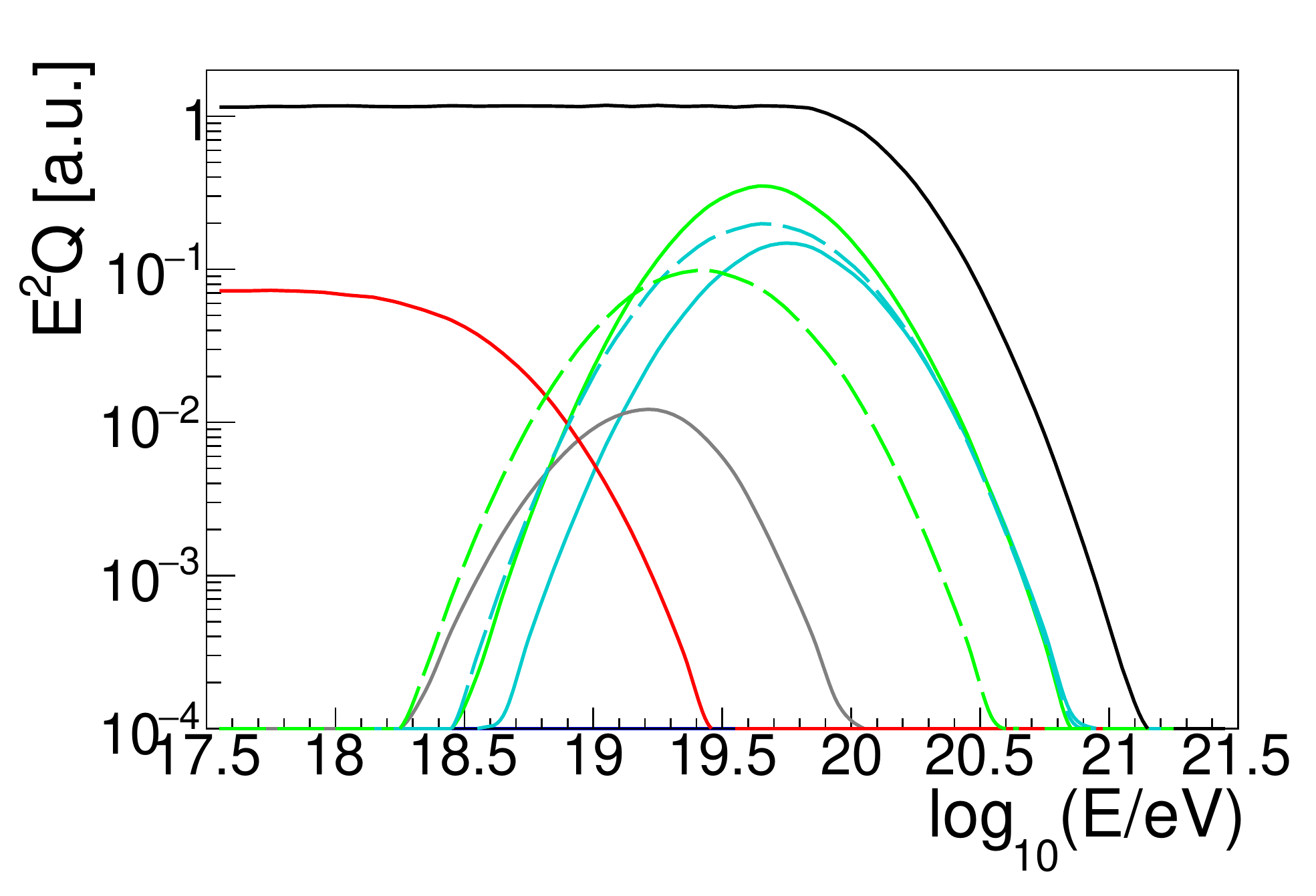}
\includegraphics[width=0.4\textwidth]{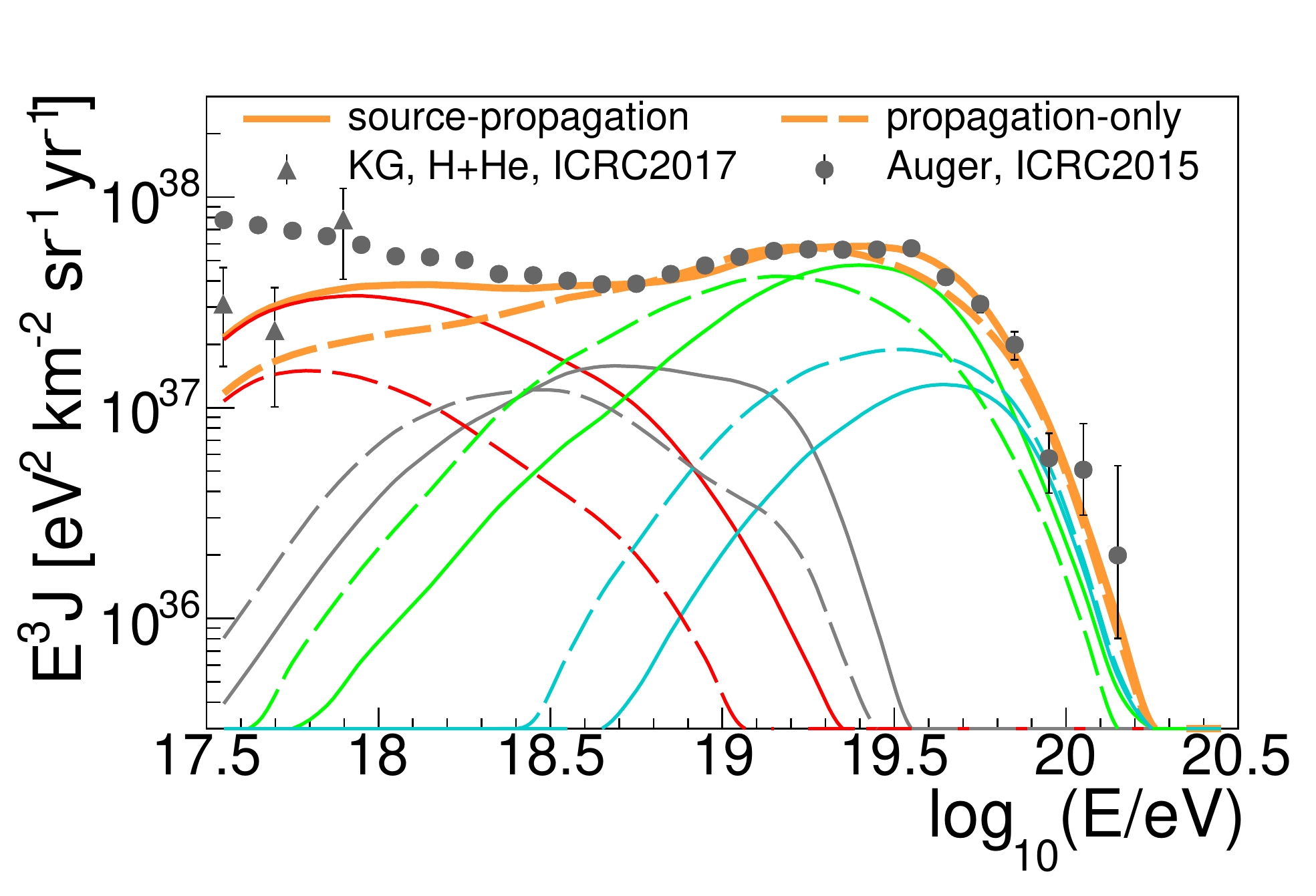}
\caption{Comparison between propagation-only model (dashed curves), corresponding to \citet{Zhang:2017moz}, and source-propagation model (solid curves), including the nuclear disintegration in the source corresponding to \citet{Biehl:2017zlw}, for the same injection composition and parameters. Left: Cosmic-ray fluxes escaping from the source (multiplied by $E^2$). Dashed curves refer to the spectra coming out of the source given by Eq.~(7) in \citet{Zhang:2017moz} (green $A=16$ and cyan $A=28$). Solid curves refer to a power law acceleration spectrum $\propto E^{-2}\exp(-E/E_{\mathrm{max}})$ (black curve, sum of $60\%$ $A=16$ and $40\%$ $A=28$) injected within the source and to the groups of isotopes generated by the interactions in the source (red $A=1$, grey $2 \leq A \leq 4$, green $5 \leq A \leq 24$, cyan $25 \leq A \leq 28$). Right: Cosmic-ray fluxes at detection (multiplied by $E^{3}$). For comparison, the Auger data points from \citet{Valino:2015} and the KASCADE-Grande data points (for the light component, H and He) from \cite{kascade:2017} are shown. The isotope groups are defined as reported above. The plots are obtained using the parameters corresponding to the best-fit as reported in \figu{bestfit}, and are independently normalized to the UHECR flux.} 
\label{fig:sourceprop}
\end{figure}

Due to their low-luminosity, LL-GRBs can be only observed in the local universe. On the other hand, LL-GRBs can have much longer durations than their high-luminosity counterpart, which is likely associated to the core-collapse supernovae progenitor scenario; they are thus assumed to exist up to high redshifts. For this reason, the propagation of the UHECRs ejected by a population of (in the cosmologically co-moving frame) identical LL-GRBs is simulated up to $z=6$, including the production of cosmogenic neutrinos. This is computed with the {\it SimProp} code \citep{Aloisio:2017iyh}, using the extragalactic background light from \citet{Gilmore:2011ks} and the TALYS photo-disintegration model in \citet{Koning:2007}, whose implementation is explained in \citet{Batista:2015mea}. 
We parametrize the evolution of the LL-GRBs with redshift relative to the star formation rate (SFR) given in \citet{Hopkins:2006bw} as $(1+z)^{m} \times H_{\mathrm{SFR}}(z)$; we consider $0\le m \le 1$ in this work.

We perform a fit of the UHECR spectrum \citep{Valino:2015} and composition \citep{Porcelli:2015}, as measured by the Pierre Auger Collaboration, in two steps. First, we fit the UHECR spectrum and composition above $10^{19}$ eV (super-ankle component). Second, in order to describe the transition to the next (sub-ankle, which can be of Galactic origin) component at lower energies, we model the end of that population as an additional power law spectrum. We then re-fit the relative weights of the sub- and super-ankle component, considering the energy range above $10^{18}$ eV. 

In total, our source-propagation model has the following parameters: the collision radius $R$ [km] (degenerate with the Lorentz factor $\Gamma$), the X-ray luminosity $L_{\mathrm{X}}$ [$\mathrm{erg\,s^{-1}}$], the emissivity of the extragalactic component $\mathcal{L_{\mathrm{ej}}}$ [$\mathrm{erg\,Mpc^{-3}\,yr^{-1}}$], the normalization of the sub-ankle component $f_{\mathrm{Gal}}$ (in terms of percentage of the total flux, and we define it at a fixed energy corresponding to $\log_{10}(E/\mathrm{eV})=17.5$), and the spectral index of the sub-ankle component $\alpha$. The quality of the fit is evaluated by computing the $\chi^2$ of the unfolded spectrum and composition data points. The latter ones are treated using the $\ln A$ parametrization given in \citet{Abreu:2013env}. The quantity $\mathcal{L_{\mathrm{ej}}}$ is referred to the total CR spectrum ejected by the source. Since we use a source-propagation model in this study, it is possible to compute this quantity and compare it  to the injection spectrum $\mathcal{L_{\mathrm{inj}}} \propto \int E \cdot E^{-2} \exp(-E/E_{\mathrm{max}}) dE$. The baryonic loading $\xi_{\mathrm{A}}$, {\it i.e.}, the ratio between energy injected as CR nuclei and the total X-ray energy $E_{\mathrm{X}}$, can be then obtained as $\xi_{\mathrm{A}} = \mathcal{L_{\mathrm{inj}}}/(\dot{n}_{\mathrm{LL-GRB}}(z=0) \cdot E_{\mathrm{X}})$. In this work, we adopt $\dot{n}_{\mathrm{LL-GRB}}(z=0)=300 \, \mathrm{Gpc^{-3}\,yr^{-1}}$, in agreement with the results of \citet{Liang:2006ci}, where $\dot{n}_{\mathrm{LL-GRB}}(z=0)=325^{+352}_{-177} \, \mathrm{Gpc^{-3}\,yr^{-1}}$ is found. Changes of the value of the local rate of the LL-GRBs and of the total duration of the burst are degenerate with the baryonic loading.

We show in Fig.~\ref{fig:sourceprop} a comparison between the propagation-only model (dashed curves), corresponding to \citet{Zhang:2017moz}, and the source-propagation model (solid curves), including the nuclear disintegration in the source corresponding to \citet{Biehl:2017zlw}, for the same injection composition and parameters. In the propagation-only model, the interactions in the source are not taken into account, and the ejected CR spectra into the extragalactic space are defined by {\em ad hoc} functions; they directly represent the injection composition. We also show for comparison the CR spectra (solid curves) obtained by using a $\propto E^{-2} \exp(-E/E_{\mathrm{max}})$ spectrum at injection (shown as black curve), including the disintegration in the source, and applying the hard escape mechanism. In the case of the source-propagation model, only one representative isotope for each isotope group is propagated. In order to compare the models, we first normalize the propagated CR fluxes to the measured spectrum and then derive the normalization of the spectra at the source. The most relevant difference  is an escaping flux of nucleons, which are generated during the development of the nuclear cascade within the source. By comparing the models after propagation (Fig.~\ref{fig:sourceprop}, right panel), a clear deficit of the light component at the lowest energies is visible in the propagation-only model compared to the source-propagation model, which could eventually require a stronger source evolution in the propagation-only case in order to describe the data. Note that  that the light component of the escaping flux has a softer spectrum compared to the other ones, because neutrons are not magnetically confined \citep{Baerwald:2013pu}. This factor turns out also to be fundamental in order to describe the CR data in the whole energy range, as already noticed in \citet{Aloisio:2013hya,Globus:2015xga,Unger:2015laa}. In addition, in a propagation-only model, the neutrino production in the source cannot be computed directly.

\section{Results and Discussion}\label{sec:results}

\begin{figure*}[t!]
\centering
\includegraphics[width=0.39\textwidth]{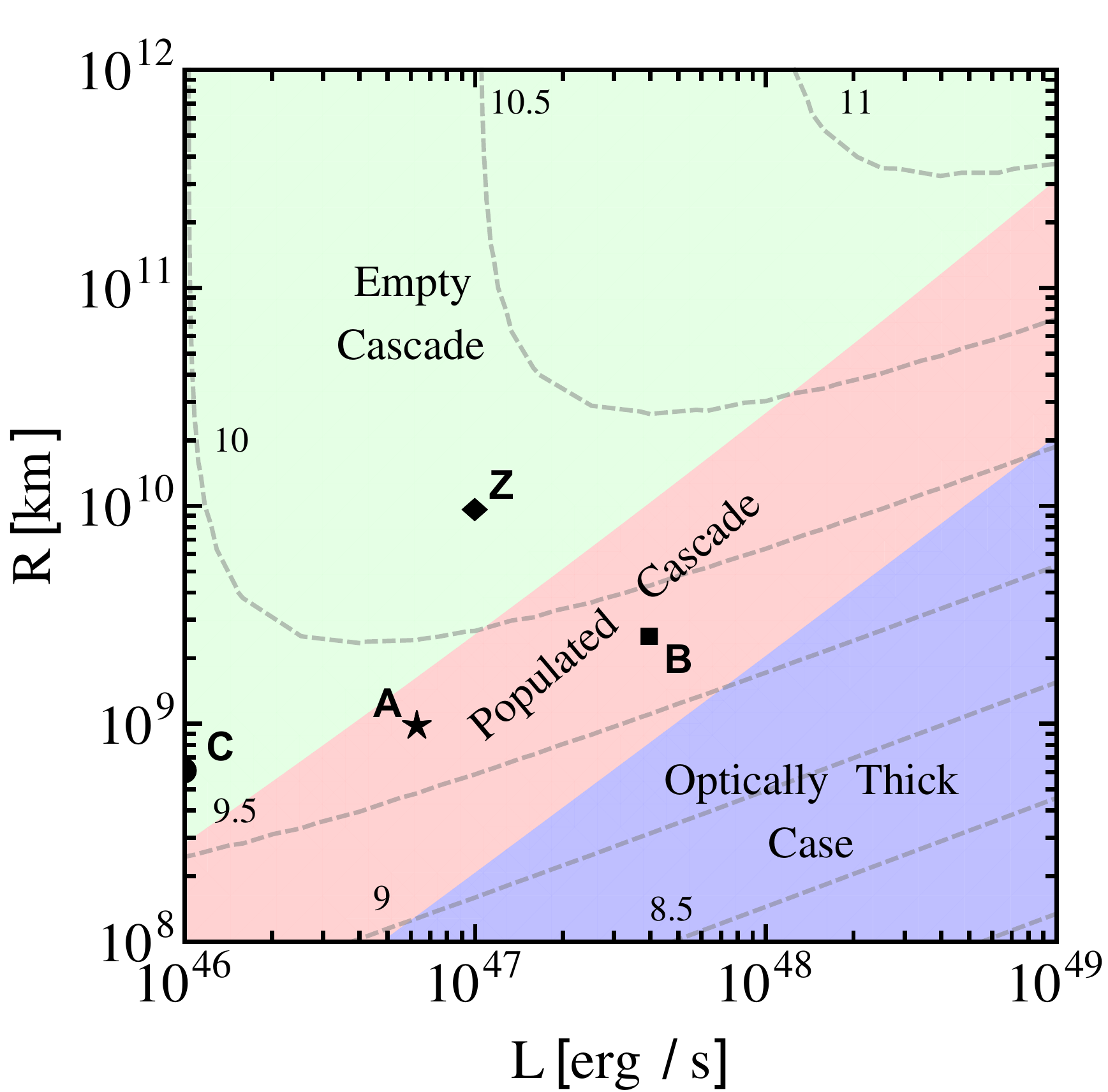}
\includegraphics[width=0.43\textwidth]{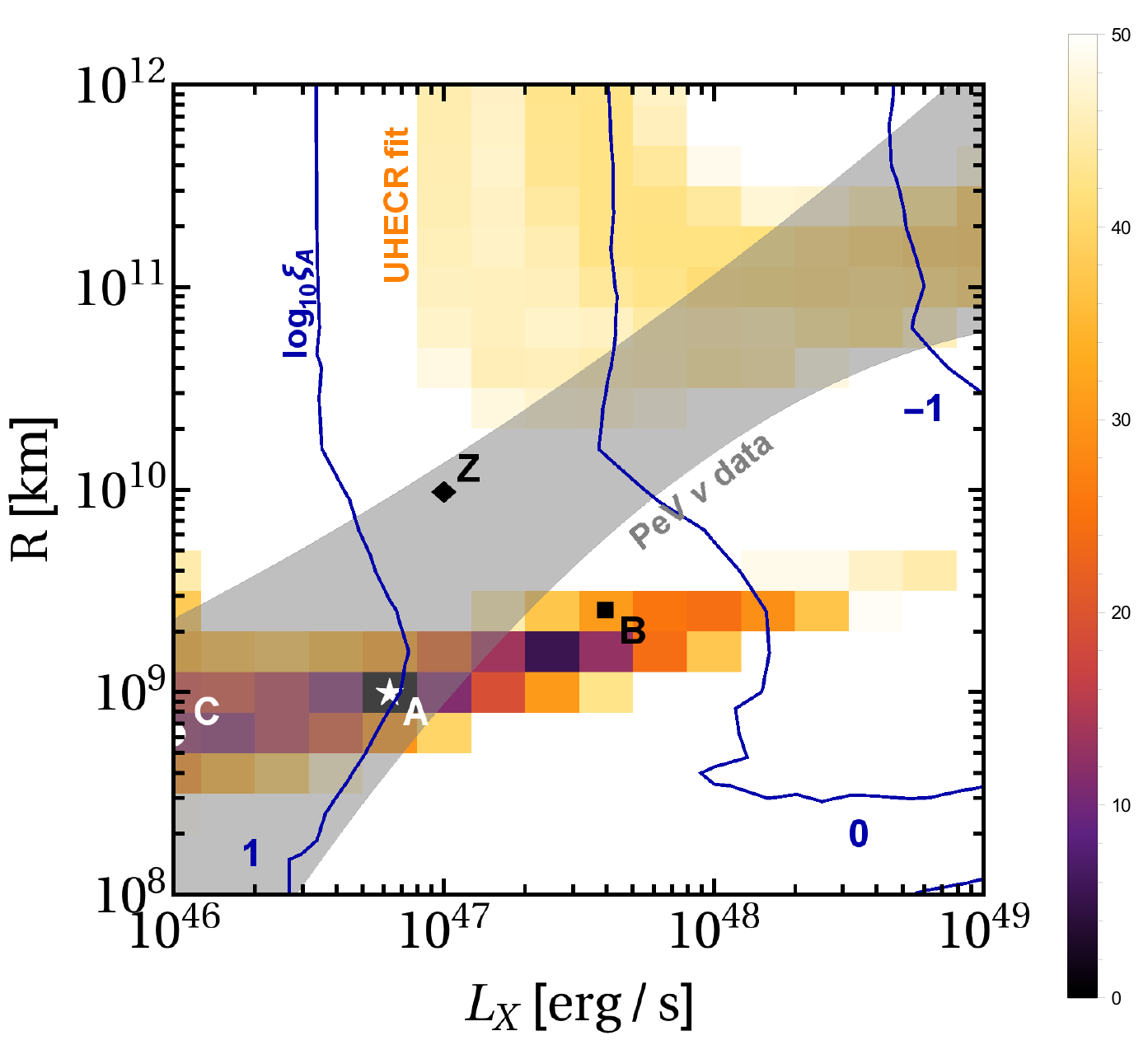} \hspace{0.4cm} 
\caption{Parameter space study as a function of X-ray luminosity $L_{\mathrm{X}}$  and collision radius $R$. {\it Left panel}: Different regimes in the parameter space for the nuclear cascade to develop in the source (shaded regions), as discussed in the main text. The curves show  $\log_{10}(E_{\mathrm{max}}/\mathrm{GeV})$, with $E_{\text{max}}$ being the obtained maximal energy for the injected isotope $A=28$ in the shock rest frame.  {\it Right panel}: Results of the fit to UHECR data (colored contours) and PeV neutrino data (gray-shaded band) as a function of $L_{\mathrm{X}}$ and $R$ (color scale: $(\chi^2-\chi^2_{\mathrm{min}})$ of the fit, gray-shaded band: neutrino PeV data including uncertainties). The blue curves show isocontours of $\log_{10} \xi_{\mathrm{A}}$ obtained from the cosmic-ray fit (and corresponding to the rate, $\dot{n}_{\mathrm{LL-GRB}}(z=0)=300 \, \mathrm{Gpc^{-3}\,yr^{-1}}$, and to the duration, $2\cdot 10^5$ s, adopted in this study; the result is degenerate in the product of these three parameters). For each point $(L_{\mathrm{X}}, R)$, the values of the other parameters that minimize the $\chi^2$ are used. In both panels, the stars indicate the parameters describing both UHECR and neutrino data (point A) and the diamond represents the parameters of the benchmark in \citet{Zhang:2017moz} (point Z). Points B and C, on the same $E_{\text{max}}$ contour as the best fit, are used as additional points for discussing the radiation density in the source (see text and Fig.~\ref{fig:ABC}).}\label{fig:fit}
\end{figure*}

\begin{figure*}[t!]
\centering
\includegraphics[width=0.4\textwidth]{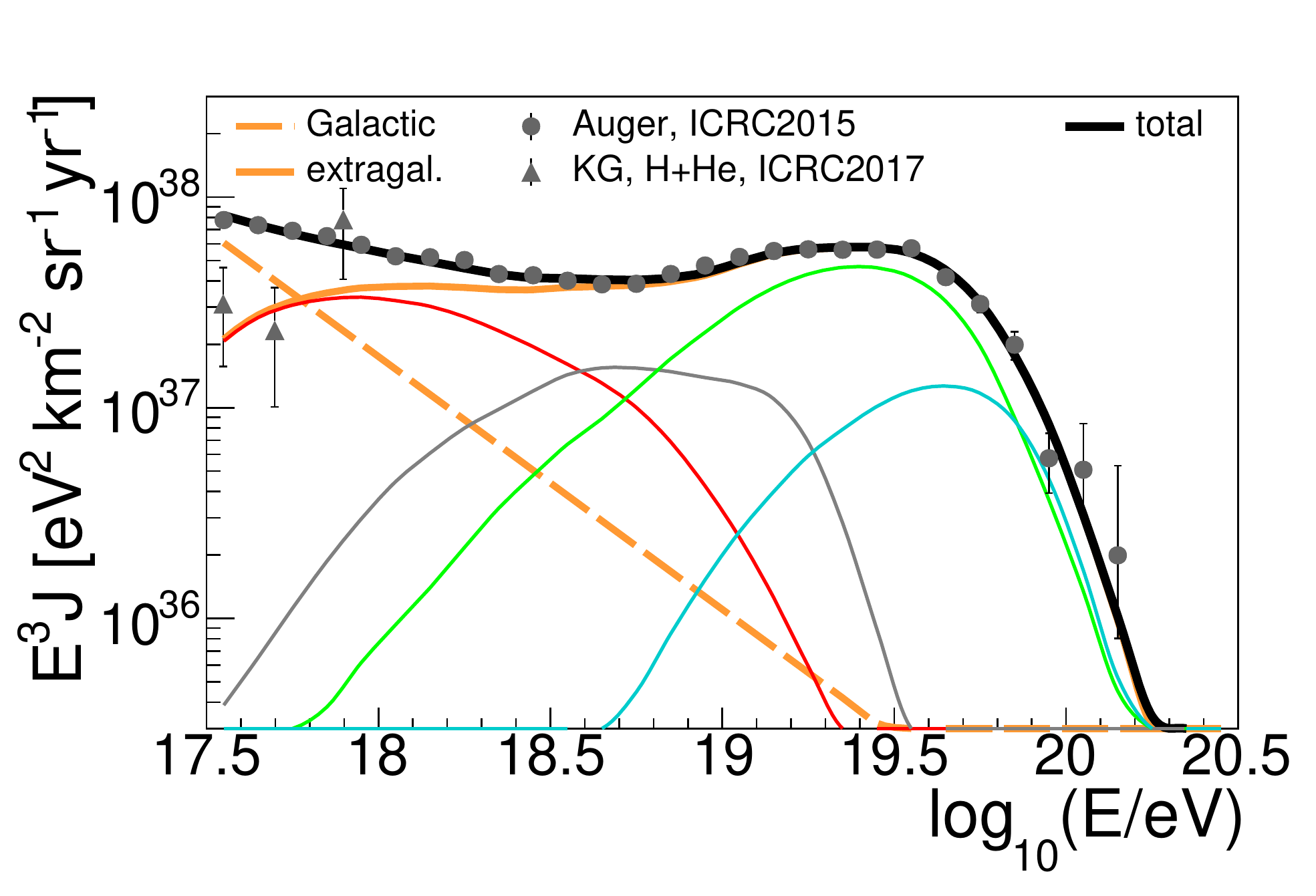}
\includegraphics[width=0.4\textwidth]{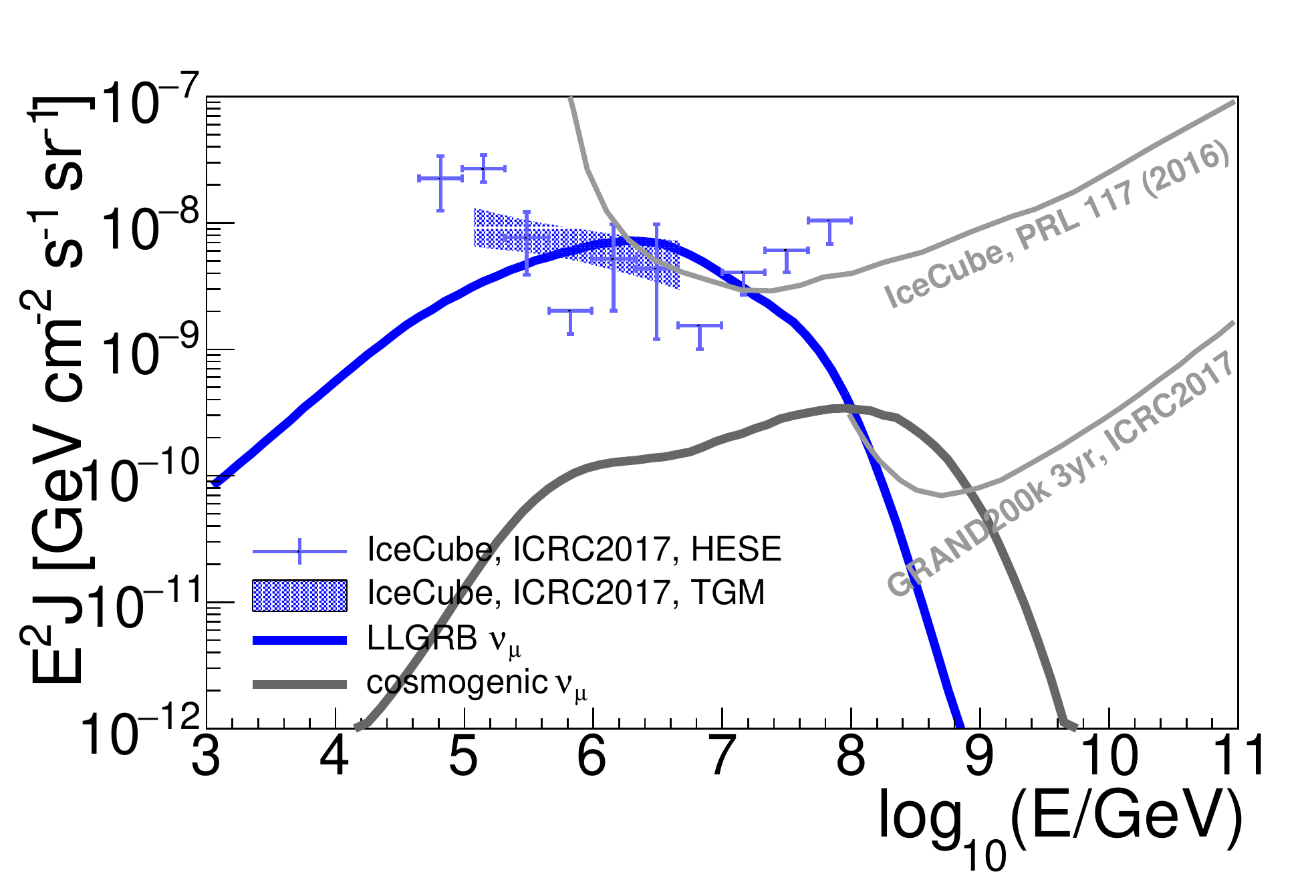}
\\
\includegraphics[width=0.83\textwidth]{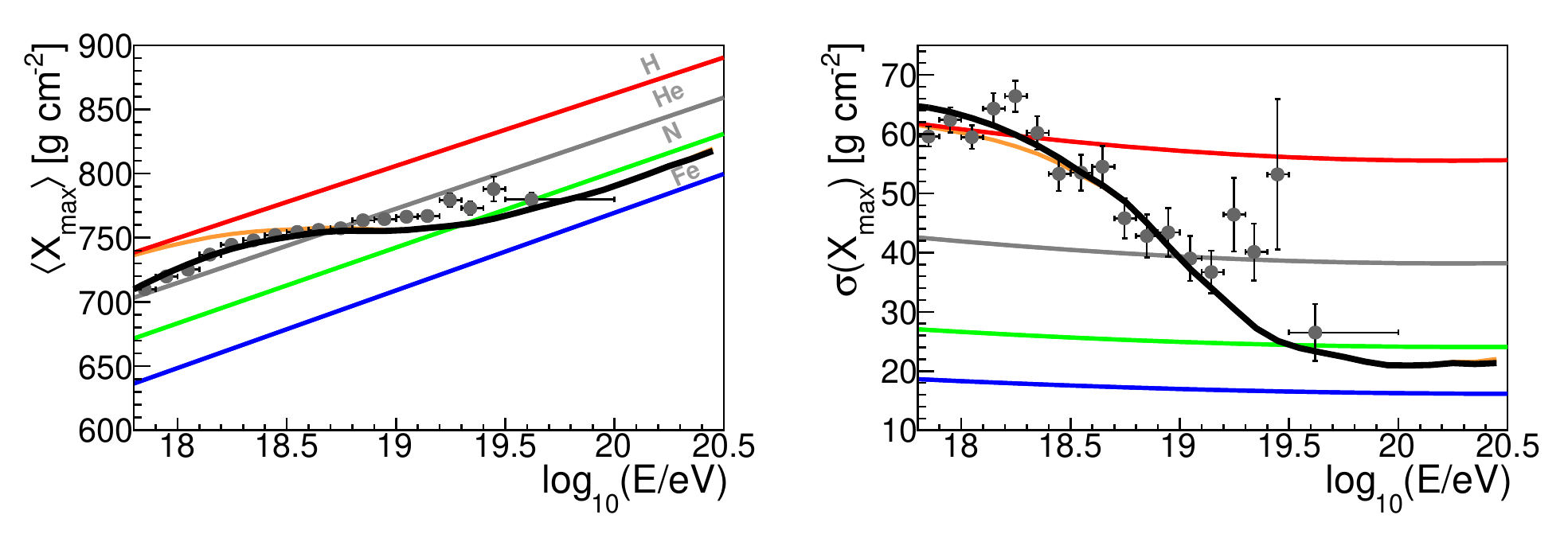}
\caption{Cosmic-ray and neutrino observables corresponding to a parameter space point describing both UHECR and neutrino data at the highest energies (denoted by the stars in \figu{fit}, $L=10^{46.8} \mathrm{erg\,s^{-1}}$, $R=10^{9}$ km, with $\xi_{\mathrm{A}} \approx 10$).
{\it Upper right panel}: Predicted muon neutrino spectrum from LL-GRBs and cosmogenic neutrinos, compared respectively to the data from the High Energy Starting Events (HESE) and the Through Going Muons (TGM) at IceCube~\citep{kopper:2017} and to the cosmogenic limits from IceCube \citep{Aartsen:2016ngq} and GRAND \citep{Fang:2017mhl}. {\it Upper left panel}: Simulated energy spectrum of UHECRs, multiplied by $E^3$; and its (extragalactic) components from (groups of) different nuclear species (thin, groups defined as in Fig.~\ref{fig:sourceprop}). The orange dashed curve represents the sub-ankle component (which may be of Galactic origin), while the solid orange curve represents the extragalactic one. For comparison, the Auger data points from \citet{Valino:2015} and the KASCADE-Grande data points (for the light component, H and He) from \cite{kascade:2017} are shown. {\it Lower panels}: Predictions (sub-ankle and extragalactic, thick black curve, and extragalactic-only, thin orange curve) and data~\citep{Porcelli:2015} on the average (left) and standard deviation (right) of the $X_{\mathrm{max}}$ distributions as a function of the energy. For predictions, EPOS-LHC~\citep{Pierog:2013ria} is assumed as the interaction model for UHECR-air interactions.}\label{fig:bestfit}
\end{figure*}

Depending on the radiation density in the source, photo-nuclear interactions can trigger the subsequent disruption of higher mass nuclei into lower mass fragments. As a consequence, the so-called nuclear cascade can develop, leading to the population of many different isotope species in the source. We show in Fig.~\ref{fig:fit} (left panel) different regimes in the parameter space for the nuclear cascade (shaded regions) as a function of X-ray luminosity $L_{\mathrm{X}}$ and collision radius $R$ for the heaviest injected mass, $A=28$. If the photon density in the source is not high enough to cause efficient disintegration, only a few species with masses close to the injected composition are populated (empty cascade). With increasing energy density, nuclei interact more efficiently with these photons such that the source becomes optically thick to photo-hadronic interactions of heavy nuclei and the nuclear cascade efficiently feeds energy into lower mass nuclei (populated cascade). For extremely high radiation densities the source becomes opaque to photo-hadronic interactions of nucleons (optically thick case), such that most of the baryonic energy is stored in protons and neutrons.
We also show the point Z corresponding to $R=10^{10}$ km and $L_{\mathrm{X}}=10^{47}$ $\mathrm{erg\,s^{-1}}$, as the representative point in the parameter space used in \citet{Zhang:2017moz}. In the right panel of Fig.~\ref{fig:fit} we show the result of the fit of the cosmic-ray spectrum and composition. The region of the parameter space, where the cosmic-ray data are best reproduced, clearly follows the contour of the maximum energy $E_{\mathrm{max}}\approx 10^{9.7}$ GeV in the source, depicted in the right panel of Fig.~\ref{fig:fit}. In the left panel of Fig.~\ref{fig:fit} we superimpose the region where the source neutrino flux is within $1\sigma$ from the IceCube PeV data points \citep{kopper:2017}. This region clearly shows that, in order to account for the IceCube flux, a moderate level of disintegration in the source is implied. We checked that all point of our parameter space are consistent with radiation constraints, i.e. efficient cosmic ray acceleration is possible \citep{Murase:2013ffa}.

The cosmic-ray and neutrino observables corresponding to the parameter space point describing both data sets are shown in Fig.~\ref{fig:bestfit}. With the same parameters describing the CR data, the neutrino flux is found to be within the expectation for the through going muons at IceCube \citep{kopper:2017}. Note that the shape of the neutrino spectrum does not perfectly describe the neutrino data points, which may be an effect of the limited statistics in neutrinos, or additional contributions to the neutrino flux, such as a Galactic component~\citep{Palladino:2018evm}. 

The required emissivity to fit the UHECR data is $\mathcal{L_{\mathrm{ej}}}=5.3\times 10^{45}\,\mathrm{erg\,Mpc^{-3}\,yr^{-1}}$, that corresponds to the injected $\mathcal{L_{\mathrm{inj}}}=4.1\times 10^{46}\,\mathrm{erg\,Mpc^{-3}\,yr^{-1}}$ (both quantities have been calculated above $10^{16}$ eV). The baryonic loading required at the best fit is $\xi_{\mathrm{A}} \sim 10$, if we take into account the local rate of LL-GRBs obtained in \citet{Liang:2006ci} and a burst duration of $2 \cdot 10^5 \, \mathrm{s}$. Interestingly, this is consistent with pioneering predictions \citep{Waxman:1997ti}, and it is substantially smaller than what found with source-propagation models taking into account HL-GRBs as UHECR and neutrino sources \citep{Baerwald:2014zga,Biehl:2017zlw}. Note,  however, that it is degenerate with the local rate and duration of the GRBs, and it may be accordingly higher for shorter durations expected from observations.

\begin{figure*}[t!]
\centering
\includegraphics[width=0.4\textwidth]{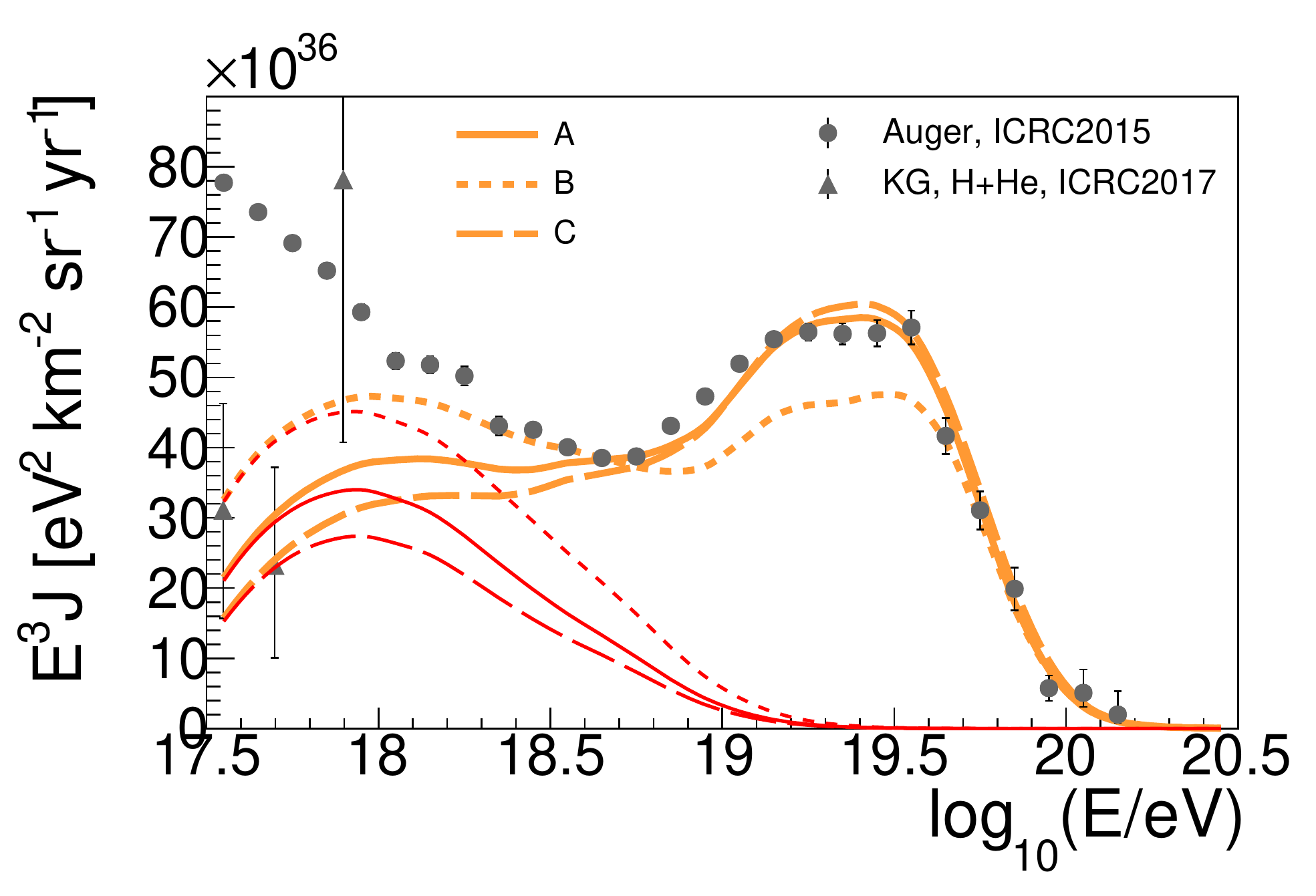}
\includegraphics[width=0.4\textwidth]{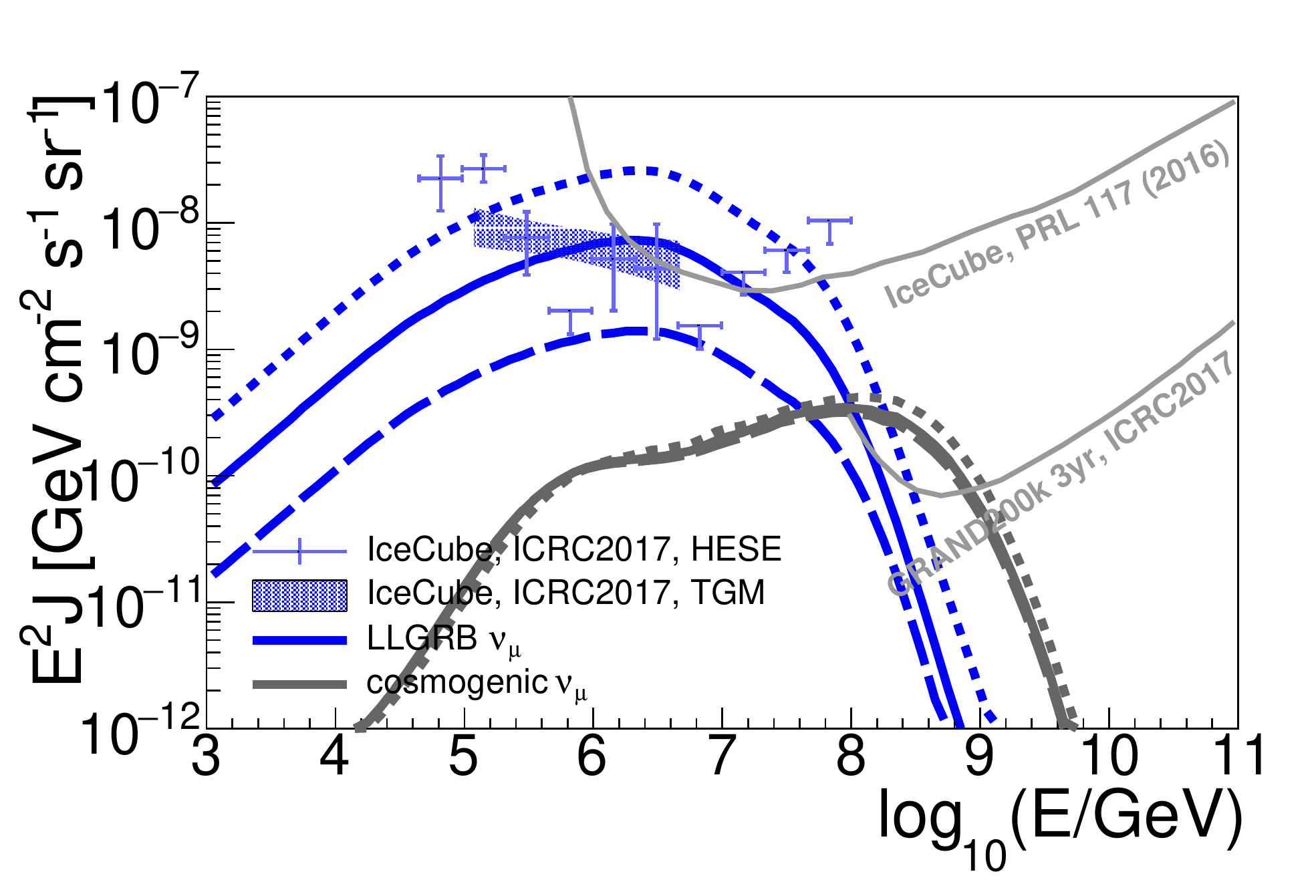}
\caption{Cosmic-ray (left panel, multiplied by $E^3$, linear scale) and neutrino (right panel, multiplied by $E^2$) fluxes at Earth corresponding to a parameter space point describing both UHECR and neutrino data at the highest energies reported in \figu{fit} (point A) and to cases with similar cutoff energy at the source, but different values for the luminosity and the radius (points B and~C), \ie, different strength of the nuclear cascade (see \figu{fit}, left panel). In the left panel, the all-particle flux (orange) is shown together with the nucleon contribution (red) of the cosmic-ray flux.}\label{fig:ABC}
\end{figure*}

In order to discuss what the effect of the radiation density in the source on the cosmic-ray and neutrino fluxes is, we show in Fig.~\ref{fig:ABC} the same observables as in the upper panel of Fig.~\ref{fig:bestfit}, for the three parameters sets marked in Fig.~\ref{fig:fit}. These parameter sets have been chosen to lie in the same maximum energy contour, so that the cosmic-ray spectra at Earth corresponding to each set exhibit similar cut-offs at the highest energies. Moving from point C to B, the enhanced radiation density in the source increases the efficiency of the interactions, producing a growing flux of light elements, which is preserved after propagation through the extragalactic space (see red lines in the left panel of Fig.~\ref{fig:ABC}). The neutrinos produced in the development of the cascade in the source, as shown in the right panel of Fig.~\ref{fig:ABC}, are strictly related to the efficiency of the disintegration in the source. The use of the source-propagation model breaks the degeneracy of the interpretation of the CR data: while both model A and B reproduce the CR spectrum above the ankle, the corresponding neutrino fluxes are clearly separated, being model B in the ``empty cascade'' region. This discrimination power is lost in the cosmogenic neutrino fluxes, due to the similar maximum energy of the parameter sets we used for this discussion.

The description of the cosmic-ray data across the ankle is a very controversial issue. Although the spectrum above EeV energies can be in principle reproduced with one-source population (as done for example in \citet{Biehl:2017zlw}), the measured composition \citep{Aab:2014kda} cannot be described by models having a prevailing light component at low energies. On the other hand, the copious production of nucleons in the interactions in the source and in the extragalactic propagation naturally grants a lighter composition while decreasing the energy. We then argue that the presence of the cutoff of the Galactic cosmic-ray population could account for a certain percentage of the CR flux at $\sim$ EeV, similarly to what has been done already, for example, in \citet{Aloisio:2013hya,Unger:2015laa,Globus:2015xga,Aab:2016zth}, and reconcile the expected composition with the measurements, as it is shown in the upper left and lower panels of Fig.~\ref{fig:bestfit}. With the introduction of this sub-ankle component, the CR spectrum can be reproduced above the EeV energies and the composition becomes heavier than what could be if only the protons produced in the propagation were considered below the ankle. Fixing the chemical composition of the sub-ankle contribution to $A=28$, the spectral index of this component is found to be $\alpha_{\mathrm{gal}}=4.2$ and the fraction of the corresponding flux at $\log_{10}(E/[\mathrm{eV}])=17.5$ is $\sim$78\%. 
The slope of the sub-ankle flux and the percentage of that with respect to the extragalactic one at $\sim$ EeV is also influenced by the source evolution, as already pointed out in \citet{Globus:2015xga}. Having investigated the effect of $m$ in the fit results, we choose here $m=1$ (closer to the GRB redshift evolution indicated in \citet{Kistler:2009mv} than the SFR, corresponding to $m=0$ in our parametrization). Although the choice of the SFR evolution with respect to a stronger one does not qualitatively affect the fit results, the nucleon flux at $\sim$ EeV is lower if $m=0$, requiring as a consequence a larger $f_{\mathrm{Gal}}$ to reproduce the sub-ankle spectrum. Another consequence is that the cosmogenic neutrino flux is expected to be lower by a factor $\sim 2$ corresponding to the best fit presented here, if SFR is used. Conversely, $m>1$ may be used to eliminate the Galactic contribution in the spectral fit; however, the flux will be still dominated by nucleons below the ankle leading to a tension with the composition data in that case.
It is also important to stress that the cosmological evolution of the LL-GRB population is yet unconstrained, due to limitations in observations. The source evolution used here is consistent with the diffusive gamma-ray background (see \cite{Globus:2017ehu}).
 
It is also interesting to notice that the cosmogenic neutrino flux is within the reach of the GRAND experiment \citep{Fang:2017mhl}. This is different from what has been found corresponding to the hypothesis of a common origin of UHECRs and neutrinos from Tidal Distruption Events (TDEs) in \citet{Biehl:2017hnb}, that are expected to have a negative evolution with redshift. However, due to the low number of detections, the evolution of the LL-GRBs with redshift is uncertain. As already pointed out for example in \citet{Heinze:2015hhp,Aab:2016zth,AlvesBatista:2018zui}, an anti-correlation between the spectral index of the ejected cosmic-ray flux and the value of $m$ exists. This is due to the fact that a positive evolution with redshift naturally softens the propagated CR flux at the lowest energies, allowing to have very hard CR spectra at the escape from the source: the escape mechanism used in this work corresponds to an effective spectrum $\propto E^{3}$. Viceversa, for a negative evolution, it is natural to expect softer CR spectra at the escape from the source, compared to what is used here ($\propto E^{-1}$, corresponding to the direct escape). While a consistent description of the UHECRs and PeV neutrino data points can be found in the two different scenarios of TDE and LL-GRB sources, a discrimination between those is given by the detectability of the cosmogenic neutrinos, which strongly depends on the redshift evolution.

Sources with dense radiation fields are usually opaque to high-energy gamma-rays as they scatter off the lower energy X-ray photons in annihilation processes. However, the target photon spectrum is only measured in a small energy band and its behaviour beyond that is uncertain. In \citet{Murase:2015xka,Biehl:2017hnb}, it has been shown that, depending on the spectral indices, the source can be optically thick to gamma-rays ranging from MeV to PeV. For the spectral indices we use in this work, gamma-rays even beyond PeV energies could be trapped. To get a rough estimate for the detection potential, we calculated the gamma-ray cascades from escaping EeV photons originating from $\pi^0$ decays. In fact, we find that high-energy photons can be expected in an energy range from MeV to TeV, which is suitable for CTA for example. 
Further investigation is needed to determine whether the expected sensitivity of CTA above 10 GeV \citep{Acharya:2017ttl} can be reached, as it depends very much on the low energy target photons.

The systematic uncertainty on the measured CR scale has not been taken into account in this work, which means that our model can reproduce the UHECR data even at the energy calibration face value. Note, however, that the energy calibration is degenerate with the acceleration efficiency, which means that solid conclusions on the shift of the energy scale or the acceleration efficiency cannot be obtained \citep{Biehl:2017hnb}. We have also tested a distribution of sources over luminosity, using the luminosity function as defined in \citet{Liang:2006ci}. Since the results are very similar to our \figu{bestfit} for an appropriate choice of the collision radius and the acceleration efficiency, we do not explicitly show them here.



\section{Summary and conclusions}\label{sec:conclusions}

We have demonstrated that a global description of the cosmic-ray and  neutrino data at the highest energies can be obtained by considering LL-GRBs as the sites of acceleration and interaction of the cosmic rays. 
We have shown that if the diffuse neutrino flux is to be powered by LL-GRBs, high photon densities in the source are required for efficient neutrino production. As a consequence, nuclei will disintegrate in the source, and the nuclear cascade developing within the source has to be taken into account. Our results are therefore based on a source-propagation model including the nuclear cascade in the source and cosmic-ray propagation. 

Interestingly, the light nuclei and nucleons (protons and neutrons) produced in the nuclear cascade can be used to describe the cosmic-ray spectrum and composition below the ankle at $5 \cdot 10^{18} \, \mathrm{eV}$. For a detailed analysis, we have included the next population dominating the cosmic-ray flux at energies $\lesssim 7 \cdot 10^{17} \, \mathrm{eV}$ as an unconstrained additional model component -- which may be of Galactic origin. As a consequence, we have obtained a near-perfect description of cosmic-ray spectrum and composition across the ankle, while at the same time powering the neutrino flux at the highest energies.

In conclusion, the efficient modeling of the processes in the jet together with the extragalactic propagation allows a direct connection between data and the characteristics of the source. 
The investigation of alternative source classes to HL-GRBs and AGN blazars is motivated by constraints on the diffuse contribution from recent IceCube stacking analyses. Therefore, alternative scenarios, including LL-GRBs, are potentially needed to describe the diffuse IceCube neutrinos. Especially if the connection between the neutrinos and the UHECRs exists, it is likely that strong enough magnetic field effects on the secondary pions, muons, and kaons break the correlation between neutrino peak energy and  maximal cosmic ray energy, as we have in LL-GRBs. For the same reason, it is  difficult to postulate the UHECR connection in AGN blazars \citep{Murase:2014foa,Rodrigues:2017fmu,Gao:2018mnu}.  Thanks to our estimate of the gamma-ray cascades from escaping EeV photons, we strongly encourage future progress in experimental studies of candidate source classes such as LL-GRBs from CTA.

\acknowledgments

We thank A. Palladino, S. Petrera, I. Sadeh and M. Unger for useful discussions. 
This work has been supported by the European Research Council (ERC) under the European Union’s Horizon 2020 research and innovation programme (Grant No. 646623).

%




\software{NeuCosmA \citep{Biehl:2017zlw,Biehl:2017hnb}, SOPHIA \citep{Mucke:1999yb}, SimProp \citep{Aloisio:2017iyh}, CRPropa 2 \citep{Kampert:2012fi}, TALYS \citep{Koning:2007}}.

\bibliographystyle{aasjournal}
\bibliography{references}



\end{document}